\documentclass[aps,prd,reprint,groupedaddress]{revtex4-2}

\usepackage{graphicx}
\usepackage{lineno}
\usepackage{amsmath}
\usepackage{multirow}

\begin{document}


\title{Training toward significance with 
the decorrelated event classifier transformer neural network}


\author{Jaebak Kim}
\email[]{jbkim@charm.physics.ucsb.edu}
\affiliation{Department of Physics, University of California, Santa Barbara, California, USA}


\date{\today}

\begin{abstract}
Experimental particle physics uses machine learning for many tasks,
where one application is to classify signal and background events.
This classification can be used to bin an analysis region to 
enhance the expected significance for a mass resonance search.
In natural language processing, one of the leading neural network architectures
is the transformer. 
In this work, an event classifier transformer is proposed to bin an analysis region,
in which the network is trained with special techniques.
The techniques developed here can enhance the significance
and reduce the correlation between the network's output and the reconstructed mass.
It is found that this trained network can perform better than
boosted decision trees and feed-forward networks.
\end{abstract}


\maketitle

\section{Introduction}

\begin{figure}
\includegraphics[width=0.25\textwidth]{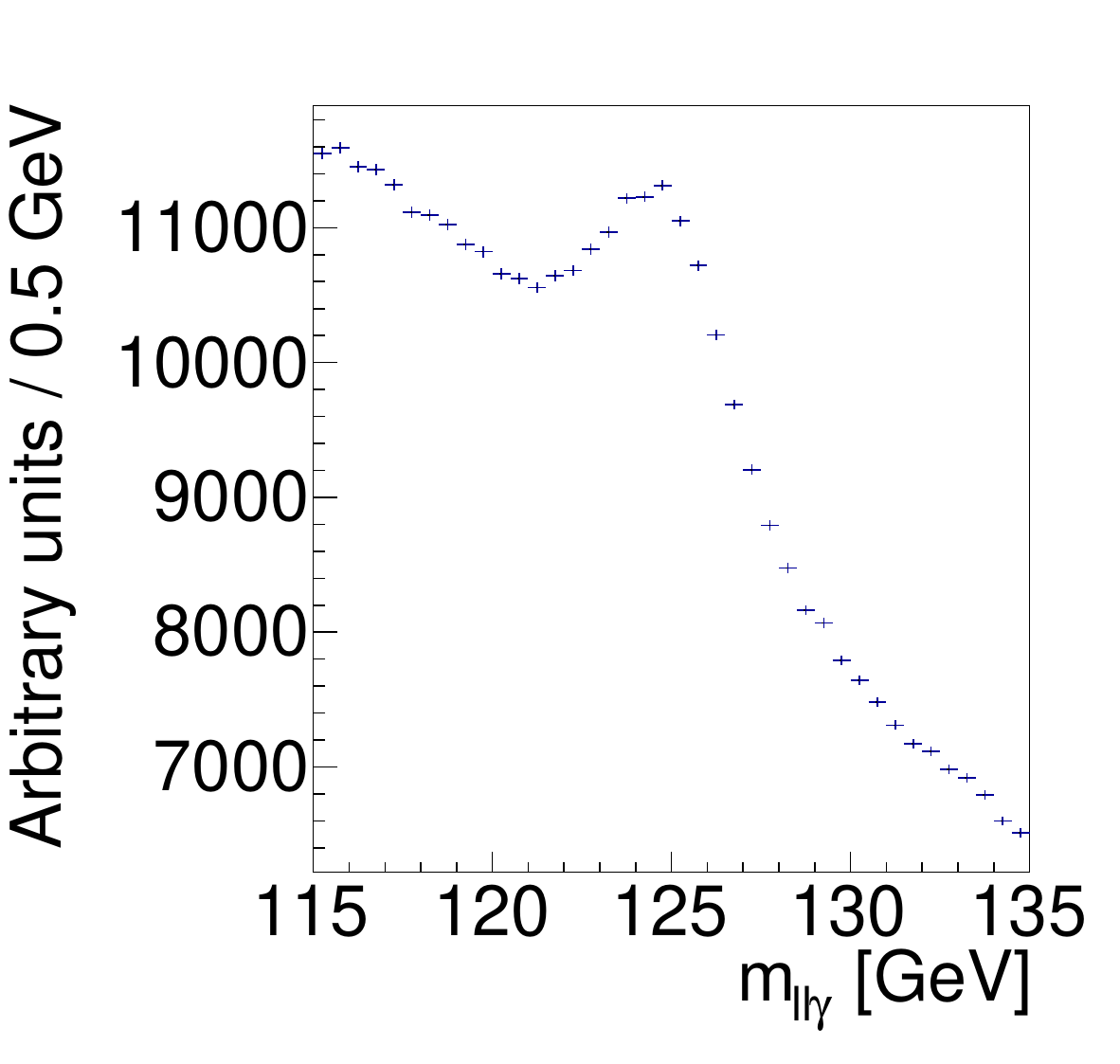}\includegraphics[width=0.25\textwidth]{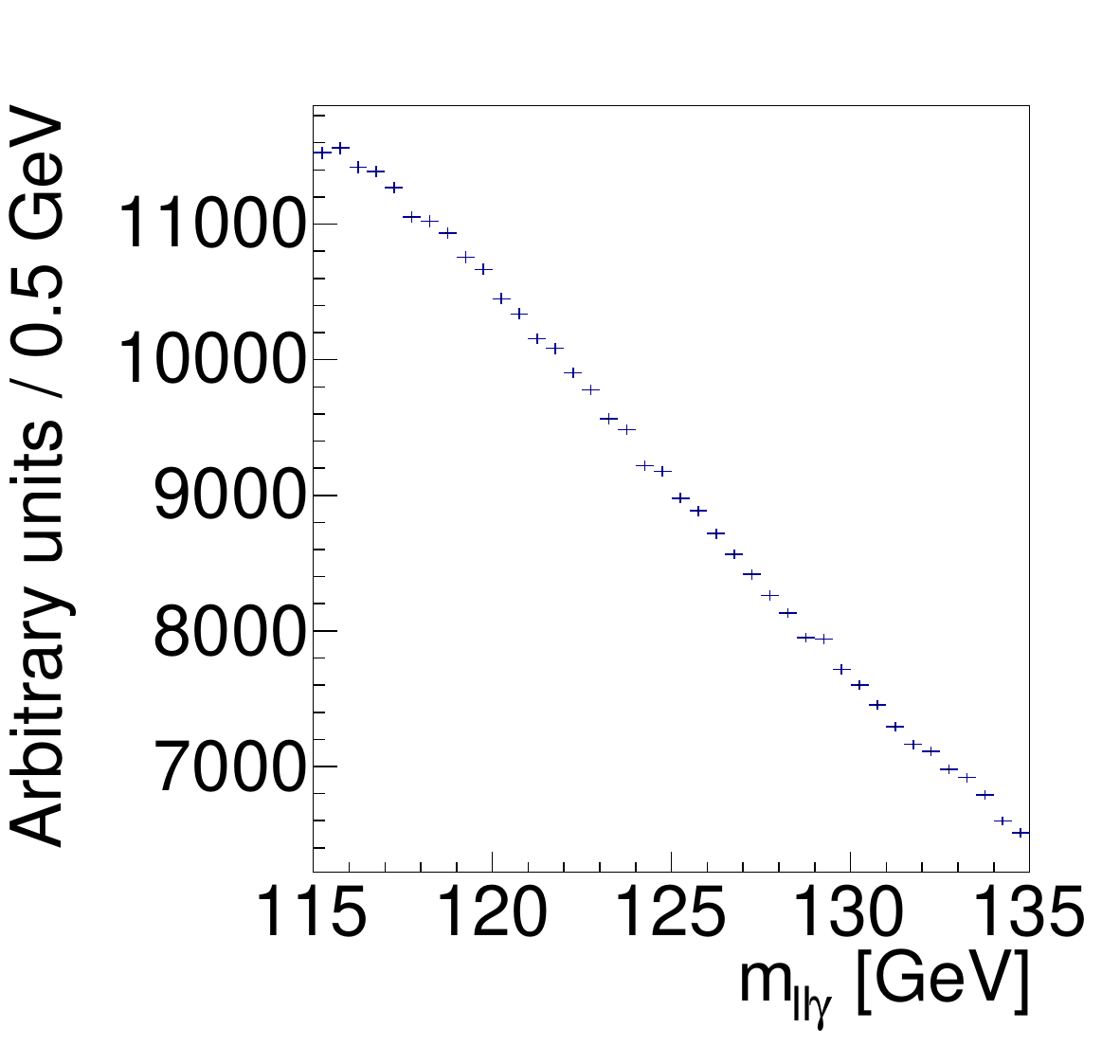}
\caption{Left: mass of reconstructed Higgs boson candidates from the $H\rightarrow Z\left(\ell^{+}\ell^{-}\right)\gamma$
decay, where a bump can be seen due to the presence of the Higgs
boson particle. The Higgs boson cross
section was scaled up by 100 to make the bump visible. 
Right: mass
of reconstructed Higgs boson candidates from the $H\rightarrow Z\left(\ell^{+}\ell^{-}\right)\gamma$
decay with the nominal Higgs boson cross section, where the bump
cannot be seen due to the background. \label{fig:MassReconParticle}}
\end{figure}

Experimental particle physics is often performed
by colliding particles
and observing their interaction with detectors. 
Particles generated from the collisions are investigated
by reconstructing the detector data. 
A common method used in particle searches
is to search for a resonance in the reconstructed mass distribution.
An example of this resonance is shown in Fig.~\ref{fig:MassReconParticle}.
Signal events create a peak that can be seen above background events.
To estimate the number of signal and background events, 
the mass distribution is fitted
with a peaking signal and a smooth background function.
The fitted mass region is set to be larger than the signal peak width
to estimate the background more precisely.
This method is referred to as ``bump hunting'' and was used
to search for the Higgs boson \cite{CMS-Higgs,ATLAS-Higgs}.
The sensitivity of the method can be quantified in
terms of significance \cite{Significance,CERN-statistics}.
A larger significance corresponds to a smaller probability 
that the bump was created by random statistical fluctuations of the background.

To increase the expected significance of the bump hunting method, requirements
are applied to create a search region to suppress the background, 
while preserving the signal.
The sensitivity of the analysis can be further enhanced by ``binning'',
where the search region is divided into multiple bins according to other variables.
However, if the binning affects 
the reconstructed mass distribution of the background,
estimating the number of signal events can become difficult.
An extreme case would be one where the background peaks similarly to the signal.
Therefore, a desirable characteristic of binning in the other variables
is that it does not affect the shape of the background distribution.
Binning can be performed using distinguishing 
features of the signal \cite{ATLAS-Higgs}, or by using machine-learning
techniques \cite{CMS-Higgs,CMS-HtoMuMu,CMS-HToZGamma} that classify the signal and background events.
A common machine-learning technique used for this task
is the boosted decision tree (BDT).

\section{Event classifier transformer neural network}


Transformer neural networks \cite{Transformer} are the leading
neural network architecture in the natural language processing field. 
The architecture has been applied to particle physics 
in a network called Particle Transformer \cite{ParticleTransformer}.
The network identifies the type of particle that produced a jet, 
which is a cluster of hadrons whose momenta lie within a cone.
In this work, a new transformer-based neural network 
called an \textit{event classifier transformer} 
is proposed that classifies signal and background events to bin a bump hunt analysis.

\begin{figure}
\includegraphics[width=0.5\textwidth]{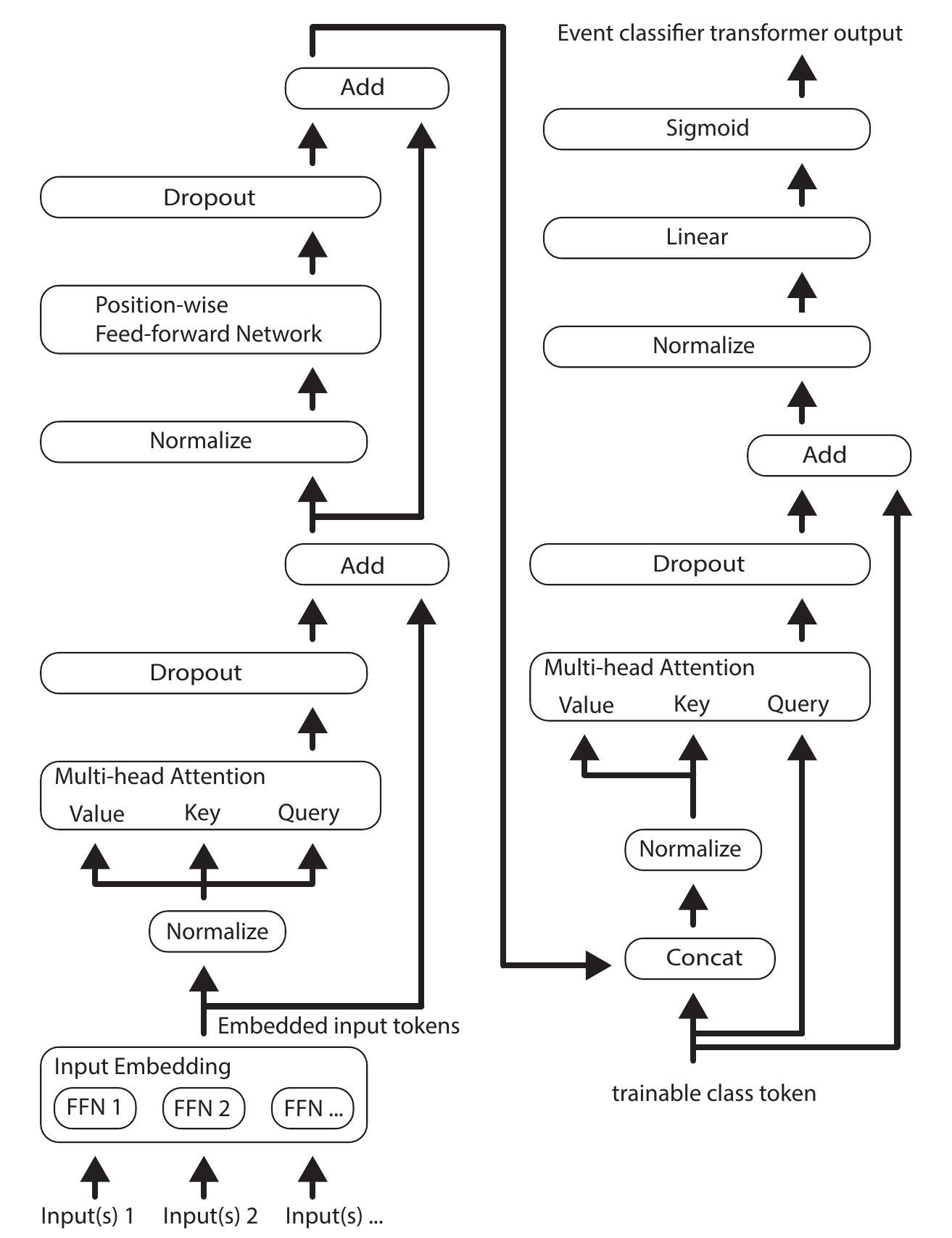}
\caption{Architecture of the \textit{event classifier transformer}.
FFN refers to a feed-forward neural network.
Normalize refers to layer normalization.
Add refers to the implementation of the residual connection.
Concat refers to a layer concatenating tokens.
Linear refers to a linear layer.
\label{fig:FeatureTransformerNetwork}}
\end{figure}

The architecture of the \textit{event classifier transformer}
is shown in Fig.~\ref{fig:FeatureTransformerNetwork}.
The inputs are the features of the event, 
such as the momenta or angles of particles. 
To be able to use the transformer architecture,
each input or each set of inputs is ``embedded'' into a token, 
which is a set of $N_{\text{repr.}}$ numbers, using separate feed-forward neural networks. 
In contrast to the transformer, there is no positional encoding 
because event features do not have sequential dependence.
The tokens are normalized using layer normalization \cite{LayerNormalization}
and passed to a multiheaded attention layer (MHA) \cite{Transformer}.
The output is summed to employ a residual connection \cite{ResidualLayer}
and normalized with layer normalization.
A positionwise feed-forward neural network \cite{Transformer}
is applied with residual connection and layer normalization,
where the same positionwise network is applied to each token.
The output is a group of tokens, 
where each token is a contextual representation of the input feature. 
Following a simplified version of the CaiT \cite{ClassToken} approach
and the Particle Transformer, 
a trainable class token is passed as a query to a MHA and
the contextualized tokens concatenated with the class token
are passed as a key and value. 
The output employs a residual connection and layer normalization and is passed to a linear layer.
The linear layer output is a single value that is passed through a sigmoid
to represent the probability that the event is from the signal process.
To prevent overtraining, dropout layers \cite{Dropout}
are used in the scores of the MHA and before the residual connection during training.
Implementation of the network is publicly available at Ref. \cite{TransformerGithub}.

\section{Training techniques for enhancing significance}

Special training techniques are developed to apply 
to the \textit{event classifier transformer} and other neural networks,
to increase the expected significance and reduce the correlation
for a bump hunt analysis.
The following new training techniques are investigated: 
\begin{itemize}
\item Specialized loss function with mass decorrelation.
\item Data scope training.
\item Significance based model selection.
\end{itemize}
These techniques are described in turn below, 
and implementation is publicly available at Ref. \cite{TransformerGithub}.

\subsection{Specialized loss function with mass decorrelation}

In a binary classification task, where the input $x$ is
provided to predict the label $y$, 
a neural network $f(x)$ can be trained 
to output a prediction $\hat{y} = f(x)$. 
In this work, signal is defined to be $y=1$
and background is defined to be $y=0$.
The network is trained with a loss function that 
is used to minimize the difference between the network's
output $\hat{y}$ and label $y$.

A common loss function is the binary cross-entropy (BCE) loss \cite{BCE},
\begin{equation}
\text{BCE}(\hat{y},y)=-y\ln(\hat{y})-(1-y)\ln(1-\hat{y}),
\end{equation}
where the minimum corresponds to the condition
\begin{linenomath}
\begin{align}
\frac{\partial\text{BCE}}{\partial\hat{y}} & = 0
\end{align}
\end{linenomath}
and is achieved when 
\begin{linenomath}
\begin{align}
\frac{\hat{y}-y}{\hat{y}\left(1-\hat{y}\right)} & = 0
\end{align}
\end{linenomath}
which implies $\hat{y}=y$.

\begin{figure}
\includegraphics[width=0.5\textwidth]{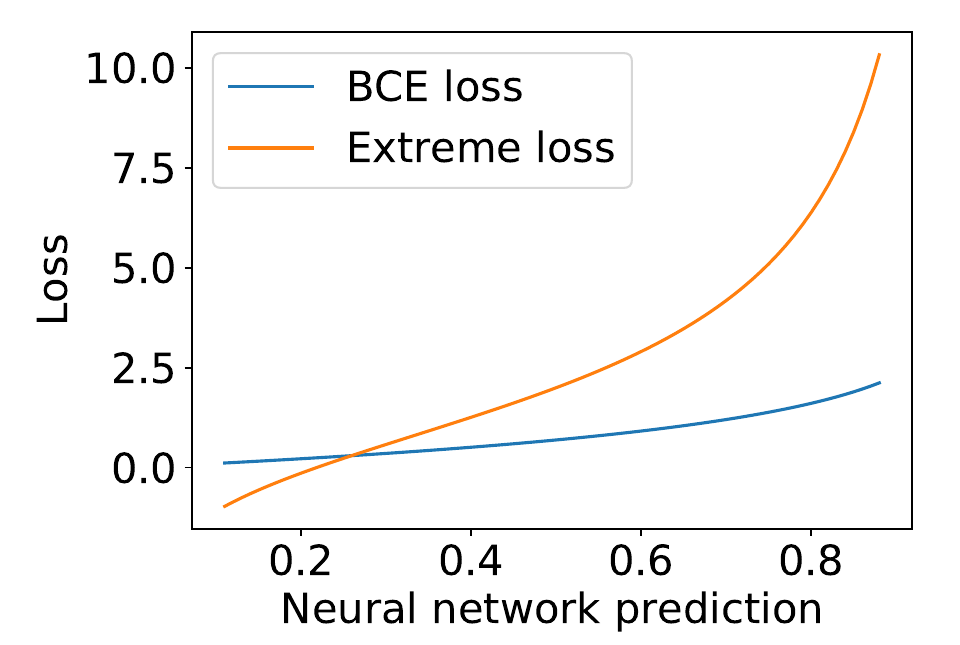}
\caption{BCE loss vs \textit{extreme} loss, when the label is $y=0$. 
\textit{Extreme} loss penalizes
the neural network more than BCE loss 
for network predictions that are close to 1. 
\label{fig:BCEvsExtreme}}
\end{figure}

In this work, to increase the penalty when $\hat{y}$ and $y$ are different,
while keeping the property that the minimum is achieved at $\hat{y}=y$,
an alternative loss inspired from BCE called \textit{extreme} loss, $E(\hat{y},y)$, is proposed:
\begin{linenomath}
\begin{eqnarray}
\nonumber E\left(\hat{y}, y\right) & = & \int\frac{\hat{y}-y}{\hat{y}^{2}\left(\hat{y}-1\right)^{2}}d\hat{y}\\
\nonumber & = & -y\left[-\frac{1}{\hat{y}}-\ln\left(1-\hat{y}\right)+\ln\left(\hat{y}\right)\right]\\
 &   & -\left(1-y\right)\left[\frac{1}{\hat{y}-1}+\ln\left(1-\hat{y}\right)-\ln\left(\hat{y}\right)\right].
\end{eqnarray}
\end{linenomath}
For a given value of $\hat{y}-y$,
the \textit{extreme} loss penalizes the neural network more 
than the BCE loss, as can be seen in Fig.~\ref{fig:BCEvsExtreme}. 
This \textit{extreme} loss heavily suppresses backgrounds that have network predictions close to 1
and also signals that have network predictions close to 0.
Because the expected significance when binning a bump hunt analysis
is sensitive to backgrounds that have high network predictions,
this loss can help to increase the significance.

To decorrelate the neural network output with the reconstructed mass, 
Distance Correlation (DisCo) regularization is used \cite{DisCo}. 
DisCo measures the dependence between $\hat{y}$ and mass, 
where the value of DisCo 
is zero if and only if $\hat{y}$ and mass are independent. 
DisCo is multiplied by a factor $\lambda$
and added to the classifier loss function:

\begin{equation}
\text{Loss}=\text{Loss}_{\text{classifier}}\left(\hat{y},y\right)+\lambda\cdot\text{DisCo}\left(\text{mass},\hat{y}\right).
\end{equation}
The DisCo term penalizes the neural network when $\hat{y}$ and mass
are correlated, where $\lambda$ sets a balance between
the neural network's performance and the degree of decorrelation.

\subsection{Data scope training \label{sec:SubsetDataTraining}}

To increase the significance 
and decorrelate the neural network output with 
the reconstructed mass for the background, 
the loss terms are calculated with different data scopes
during training:

\begin{linenomath}
\begin{eqnarray}
\nonumber \text{Loss} & = & \text{Loss}_{\text{\text{classifier}}}\left(\hat{y},y\right)\\
\nonumber & & y=\left[0,1\right]\\
\nonumber & & \hat{y}\in\text{narrow mass window}\\
\nonumber & & \\
& & + \lambda\cdot\text{DisCo}\left(\text{mass},\hat{y}\right).\\
\nonumber & & y=0\\
\nonumber & & \hat{y},\text{mass}\in\text{wide mass window}
\end{eqnarray}
\end{linenomath}


The classifier loss, such as BCE loss, is calculated in the scope
of signal and background events that fall in the narrow mass window, 
where the majority of the signal lies. 
This narrow scope decorrelates the network's output with the mass and can improve the significance.
The DisCo term is calculated in the scope of background events
that fall in the wide mass window, 
where the mass window is used in estimating the amount of background
in the bump hunt method.

\subsection{Significance-based model selection \label{sec:Significance}}

When training networks, the best-trained network among
the training epochs can be chosen based on an evaluation metric.
The loss is often used as the evaluation metric, 
but the network that has the minimum loss can be different 
from the network that has the best significance \cite{Significance,LossVsSignificance}. 
In this work, the expected significance is used as a metric 
to select the best network among the epochs.

The significance is calculated with the following steps:
\begin{enumerate}
\item Divide the dataset into bins based on the neural network's output.
The bins are constructed to have an equal number of signal events.
\item Calculate the significance of each bin.
\item Combine the significances of the bins.
\end{enumerate}

The significance \cite{Significance} of each bin is calculated with
\begin{equation}
\text{Significance}=\sqrt{2\left[\left(N_{S}+N_{B}\right)\ln\left(1+\frac{N_{S}}{N_{B}}\right)-N_{S}\right]},
\end{equation}
where $N_{S}$ is the number of signal events and $N_{B}$ is the
number of background events within a mass window containing the
5th to 95th percentile of the signal. 
Note that the mass window width can change depending on the bin. 
The combination of significance \cite{CombinePvalues} over the bins is calculated with
\begin{equation}
\text{Total significance}=\sqrt{\sum_{i}^{n}\left(\text{Significance}_{i}\right)^{2}},
\end{equation}
where $n$ is the number of bins and $i$ is the index of the bin.

\section{Example analysis}

In this work, a search for the process $H\rightarrow Z\left(\ell^{+}\ell^{-}\right)\gamma$
is considered, where $\ell^{+}\ell^{-}$ represents an $e^+e^-$ or $\mu^+\mu^-$ pair.
Such a study has been performed by CMS \cite{CMS} and ATLAS \cite{ATLAS} 
with the Run 2 LHC data using a luminosity of around $150~\text{fb}^{-1}$,
where the BDT technique was used in binning the search region.
The expected significance for this standard model process of each analysis 
is $1.2\sigma$ \cite{CMS-HToZGamma,ATLAS-HToZGamma}.
When adding the expected Run 3 LHC data \cite{LHCRun3}, 
where $250~\text{fb}^{-1}$ of data could be collected, 
and assuming similar analysis sensitivity 
the expected significance is $2\sigma$.
To reach a $3\sigma$ significance, an additional
$600\ \text{fb}^{-1}$ of data is required.
This will be achieved in the
High Luminosity LHC \cite{HL-LHC} era that is planned to start in
2029, where around $300\ \text{fb}^{-1}/\text{year}$
of data is expected. To reduce the amount of time to
obtain evidence $\left(3\sigma\right)$ of this decay, a neural network
approach is explored.

The simplified $H\rightarrow Z\left(\ell^{+}\ell^{-}\right)\gamma$ analysis
in this work searches for a resonance in the reconstructed mass distribution of the
Higgs boson candidates. To increase the significance, the search region
of the analysis is binned with specially trained neural
networks and the performance is compared with that of boosted decision trees.

The following sections describe the dataset, 
input features for the machine-learning techniques, the machine-learning techniques themselves,
evaluation metrics, experiment, and results.

\subsection{Dataset}

The dataset is generated with the Monte Carlo event generator 
\texttt{MADGRAPH5\_aMC@NLO} \cite{Madgraph5.1,Madgraph5.2}, 
particle simulator \texttt{PYTHIA8} \cite{Pythia}, 
and detector simulator \texttt{DELPHES3} \cite{Delphes},
where jet clustering is performed by the \texttt{FastJet} \cite{FastJet} package. 
For the event generation of the signal, 
the Higgs boson $\left(pp\rightarrow H\right)$
is generated with \texttt{MADGRAPH5\_aMC@NLO} 
using the Higgs Effective Field Theory (HEFT) model \cite{HEFT} 
and decayed to $H\rightarrow Z\left(\ell^{+}\ell^{-}\right)\gamma$
using \texttt{PYTHIA8}. 
The background $pp\rightarrow Z\left(\ell^{+}\ell^{-}\right)\gamma$ is generated with
\texttt{MADGRAPH5\_aMC@NLO} at leading order. 
For detector simulation, the CMS detector settings of \texttt{DELPHES3} are used.
To have samples for training, validation, and testing, a dataset
of 45 million events are generated for both the signal and
the background, for a total of 90 million events. 
The total number of signal events for each of the
training, validation, and testing datasets is scaled to have the standard model cross
section of $7.52\times10^{-3}\ \text{pb}$ with a luminosity of $\text{138\ \text{fb}}^{-1}$, 
corresponding to the Run 2 luminosity of CMS.
The background is scaled to have the standard model cross section of $55.5\ \text{pb}$
with the same luminosity as the signal.

The following requirements, which are similar to the CMS analysis \cite{CMS-HToZGamma}, 
are applied to the dataset:
\begin{itemize}
\item Trigger threshold requirements: $p_{T}^{\text{leading }\ell}>25\text{ GeV}$, $p_{T}^{\text{subleading }\ell}>15\ \text{GeV}$.
\item Background suppression requirements: $p_{T}^{\gamma}/m_{\ell\ell\gamma}>15/110$, 
minimum $\Delta R\left(\ell^{\pm},\gamma\right)>0.4$, $m_{\ell\ell}>50\ \text{GeV}$, 
$m_{\ell\ell\gamma}+m_{\ell\ell}>185\ \text{GeV}$.
\item Mass window requirement: $100<m_{\ell\ell\gamma}<180\ \text{GeV}$.
\end{itemize}
After applying these requirements, 
there are 9 million events for the signal 
and 5 million events for the background.
The large sample helps to reduce overtraining,
where the significance metric is sensitive to the sample region that has high classifier scores
and a low number of background events.

\subsection{Input features for machine-learning techniques}

The following features are used as inputs for the machine-learning techniques that bin the analysis:
\begin{itemize}

\item $\eta_{\gamma}$, $\eta_{\ell}^{\text{leading }\ell}$, $\eta_{\ell}^{\text{subleading }\ell}$:
Pseudorapidity angle of the photon and leptons.
\item Minimum $\Delta R\left(\ell^{\pm},\gamma\right)$, maximum $\Delta R\left(\ell^{\pm},\gamma\right)$:
Minimum and maximum $\Delta R$ between leptons and the photon.
\item Flavor of $\ell$: Flavor of lepton used to reconstruct the $Z$
boson, either being an electron or a muon.
\item $p_{T}^{\ell\ell\gamma}/m_{\ell\ell\gamma}$: $p_{T}$ of the reconstructed
Higgs boson candidate divided by the mass.
\item $p_{Tt}^{\ell\ell\gamma}$: Projection of the reconstructed
Higgs boson $p_{T}$ to the dilepton thrust axis \cite{Ptt}.
\item $\sigma_{m}^{\ell\ell\gamma}$: Mass reconstruction error of the $\ell\ell\gamma$
candidate estimated by binning signal events in $\eta$ and $p_{T}$
for the photon and leptons, and measuring
the signal's mass width for each bin.
\item $\cos\Theta$, $\cos\theta$, $\phi$:
Production and decay angles that determine the differential cross
section of $H\rightarrow Z\left(\ell^{+}\ell^{-}\right)\gamma$ and
$qq\rightarrow Z\left(\ell^{+}\ell^{-}\right)\gamma$ \cite{ProductionDecay1,ProductionDecay2}.
\item $p_{T}^{\gamma}/m_{\ell\ell\gamma}$, $p_{T}^{\text{leading }\ell}$,
$p_{T}^{\text{subleading }\ell}$: Momenta of the photon and leptons. 

\end{itemize}

\begin{figure}
\includegraphics[width=0.45\textwidth]{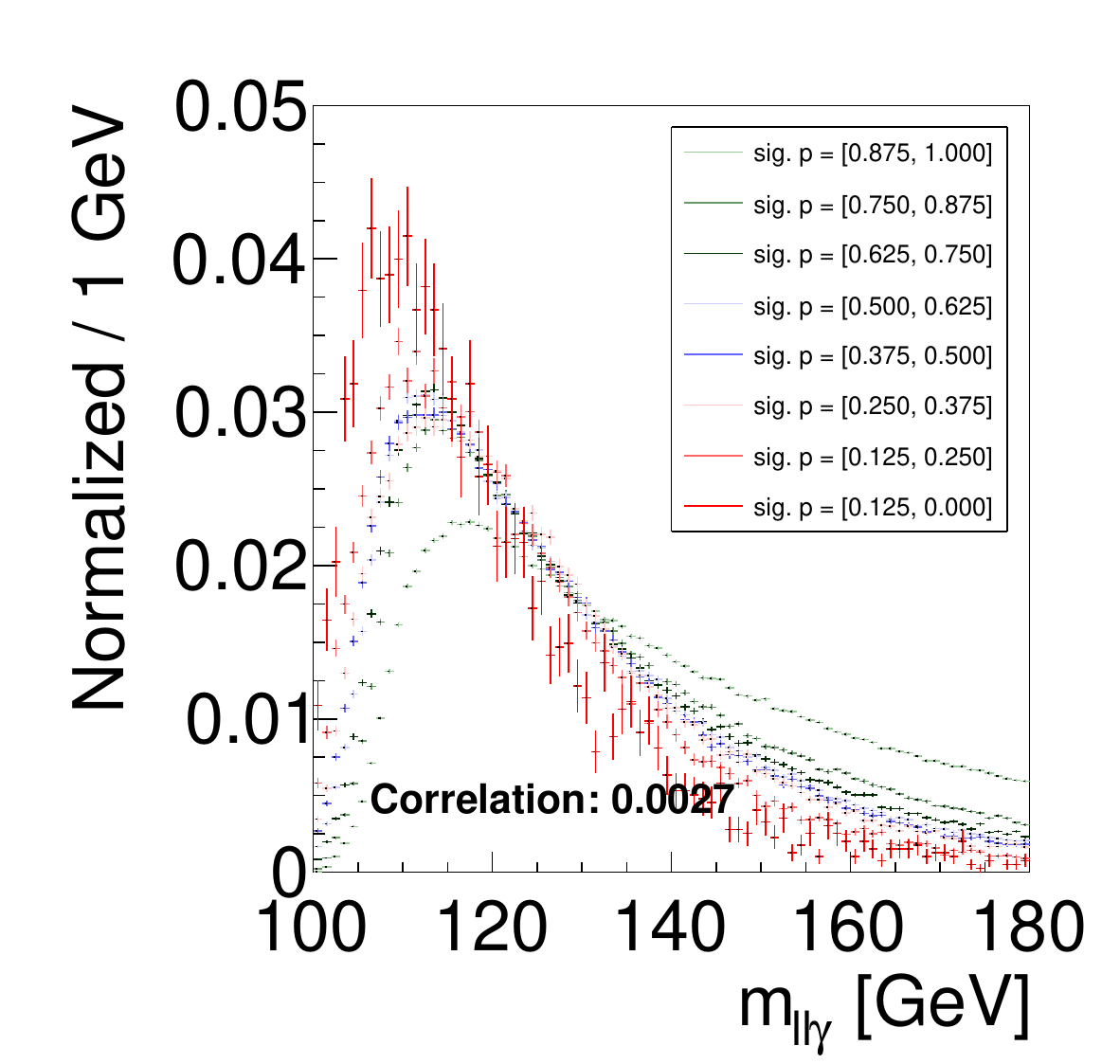}\\
\includegraphics[width=0.45\textwidth]{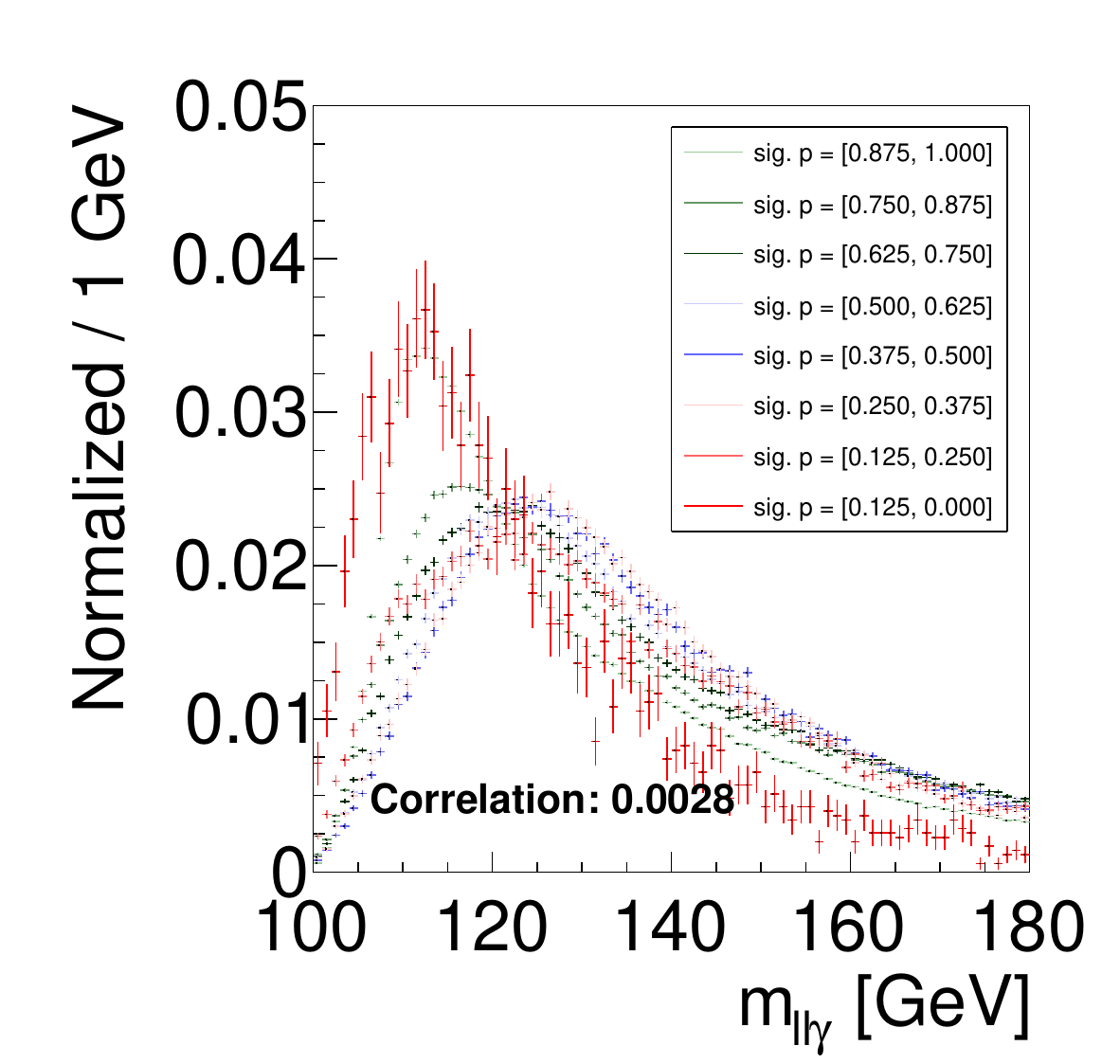}
\caption{Top: reconstructed $m_{\ell\ell\gamma}$ background distributions,
where each histogram is a bin in the XGBoost output distribution with
an equal number of signal events. 
Lower signal percentile (sig. p) values correspond to higher output values.
$p_{T}^{\gamma}/m_{\ell\ell\gamma}$,
$p_{T}^{\text{leading }\ell}$, $p_{T}^{\text{subleading }\ell}$
are not used in the training. Bottom: reconstructed $m_{\ell\ell\gamma}$
background distributions, when training includes $p_{T}^{\gamma}/m_{\ell\ell\gamma}$,
$p_{T}^{\text{leading }\ell}$, and $p_{T}^{\text{subleading }\ell}$ inputs for XGBoost. 
For certain bins, the background peaks close
to the Higgs boson mass of 125~GeV, which introduces difficulties
in estimating the number of signal events. 
Correlation represents the magnitude of difference in the shapes between the machine-learning bins.
\label{fig:XGBoost15Var}}
\end{figure}

Many of the features are correlated with the reconstructed Higgs mass.
Therefore depending on which features are included in the machine-learning technique
and which features the machine-learning technique prioritizes,
the output of machine-learning classifier can also be correlated with the mass.
Especially when including the momenta of the photon and lepton
in certain machine-learning techniques,
such as BDTs, to bin the search region, 
the background mass distribution 
in certain bins tends to peak close to the Higgs boson mass, 
which can be seen in Fig.~\ref{fig:XGBoost15Var}. 
This behavior introduces difficulties in estimating the number of signal events, so
that these features are typically excluded for the inputs of the
machine-learning technique. 
However, these features can be used as inputs
for neural networks that are trained to be
decorrelated with the reconstructed mass.

\subsection{Machine-learning techniques}

The following machine-learning techniques are compared:
\begin{itemize}
\item Boosted Decision Trees using the TMVA framework \cite{TMVA}, which
has 850 trees, with a minimum node size of 2.5\%.

\item XGBoost \cite{XGBoost} with 100 boosting rounds with a
maximum depth of 3. The BDT was optimized by modifying the number of boosting rounds,
where BDTs with more than 100 rounds have higher overtraining, 
while other performance metrics are similar.

\item Feed-forward neural network 
that has an input layer with $N$ nodes corresponding to the number of input features, 
a hidden layer with $4N$ nodes with a $\tanh$ activation function,
and an output layer with one node with a sigmoid activation function. 
The network is implemented with \texttt{PyTorch} \cite{Pytorch}.

\item Deep feed-forward neural network 
that has an input layer with $N$ nodes corresponding to the number of input features
and three hidden layers with $3N$, $9N$, and $3N$ nodes 
with a $\tanh$ activation function.
The last hidden layer is connected to a dropout layer with a rate of 0.1.
The output layer has a single node with a sigmoid activation function.
The network is implemented with \texttt{PyTorch}.
The network was optimized by modifying the dropout rate,
where the network with no dropout had significant overtraining and poorer performance,
while the network with a dropout rate of 0.2 had similar performance to the dropout 0.1 network.

\item \textit{Event classifier transformer} with $N_{\text{repr.}}=16$.
Each event feature is embedded by a separate feed-forward network that has
an input layer with one node, 
a hidden layer with 4 nodes with a GELU activation function,
and an output layer with 16 nodes.
The MHA has 4 heads, and 
the positionwise feed-forward network has
an input layer with 16 nodes,
a hidden layer with 64 nodes with a GELU activation function,
and an output layer with 16 nodes.
The dropout layer has a rate of 0.1.
The network is implemented with \texttt{PyTorch}.
The network was optimized by modifying the dropout rate, 
where the network with no dropout showed overtraining and poorer performance,
while the network with a dropout rate of 0.2 had slightly worse performance
than the dropout 0.1 network.

\end{itemize}
When training the machine-learning techniques, the signal and background samples
are weighted to have the same number of events.
All input features are normalized for neural networks.

For BDTs and networks that are trained without DisCo loss,
12 input features are used: $\eta_{\gamma}$, $\eta_{\ell}^{\text{leading }\ell}$,
$\eta_{\ell}^{\text{subleading }\ell}$, minimum $\Delta R\left(\ell^{\pm},\gamma\right)$,
maximum $\Delta R\left(\ell^{\pm},\gamma\right)$, flavor of $\ell$,
$p_{T}^{\ell\ell\gamma}/m_{\ell\ell\gamma}$, $p_{Tt}^{\ell\ell\gamma}$,
$\sigma_{m}^{\ell\ell\gamma}$, $\cos\Theta$, $\cos\theta$, and $\phi$.
These input features have minimal correlation with the reconstructed $m_{\ell\ell\gamma}$. 
The training is performed with events that have a reconstructed Higgs boson mass
range between 120 and 130 GeV.

When networks are trained with DisCo loss,
three additional features are used: $p_{T}^{\gamma}/m_{\ell\ell\gamma}$,
$p_{T}^{\text{leading }\ell}$, and $p_{T}^{\text{subleading }\ell}$.
The \textit{data scope training} method described in Sec. \ref{sec:SubsetDataTraining} is used.
The narrow mass window is from 120 to 130~GeV, and the
wide mass window is from 100 to 180~GeV.

Different combinations of loss functions are explored:
\begin{enumerate}
\item BCE loss.
\item \textit{Extreme} loss.
\item BCE + DisCo loss.
\item \textit{Extreme} + DisCo loss.
\end{enumerate}
To increase the stability of the calculation, 
the \textit{extreme} loss implementation clamps the neural network output $\hat{y}$
to a range from 0.001 to 0.999.
The $\lambda$ DisCo factor is 10 for the BCE + DisCo loss 
and 50 for the \textit{extreme} + DisCo loss
to have similar background correlations among networks trained with DisCo.

Each neural network is trained for 1200 training epochs with the Adam
optimization algorithm \cite{Adam} with a learning rate of $10^{-3}$
and a batch size of 8192.
The best model between the epochs is selected
by finding the \textit{model with highest significance} on the validation dataset.
To reduce the amount of time in training, the model is
evaluated on the validation dataset in five training epochs intervals for the
first 50 epochs, and then ten training epoch intervals for
the remaining training epochs. 
This method allows the best model search to be sensitive to the early epochs,
where the evaluation metric can change dynamically.

\subsection{Machine-learning technique evaluation metrics \label{Sec:EvaluationMetrics}}

The machine-learning techniques are evaluated with the following metrics:
\begin{itemize}
\item Expected significance.
\item Area Under the Curve (AUC) of the Receiver Operating Characteristic
(ROC) curve.
\item Correlation of $m_{\ell\ell\gamma}$ with the machine-learning technique
output.
\item Epoch of the best-trained model
\end{itemize}
The significance is calculated using the method described in Sec.
\ref{sec:Significance}, where the sample is divided into eight machine
learning technique bins with an equal number of signal events in each
bin.

The AUC curve is calculated using the \texttt{scikit-learn} \cite{Sklearn}
package with negative weights set to zero and evaluated in the Higgs
boson mass window from 120 to 130~GeV.

The correlation is measured with the following steps:
\begin{enumerate}
\item The search region is binned according to the machine-learning technique
output.
\item Normalized $m_{\ell\ell\gamma}$ histograms are created for each machine-learning technique bin.
\item The differences between the machine-learning technique bins
are measured with the standard deviation of the normalized yield for each mass bin.
\item The mean of the standard deviations is calculated.
\end{enumerate}
When calculating the correlation, the sample is divided into eight machine-learning 
technique bins with an equal number of signal events in each
bin.

\subsection{Experiment and results}

A repeated random subsampling validation procedure \cite{RandomSubsample}
is used to evaluate the machine-learning techniques, 
where three trials are done.
The number of trials is limited due to the long training time of the neural networks.
The validation procedure follows the steps below:
\begin{enumerate}
\item The events in the dataset are randomly shuffled 
and divided equally for the training, validation, and test datasets.

\item The machine-learning technique is trained using the training dataset.
For neural networks, the best model between the epochs 
is selected using the expected significance metric on the validation dataset.

\item The performance of the machine-learning technique using the evaluation
metrics is evaluated on the test dataset.

\item The steps above are repeated $N$ times, where in this work $N=3$.
The machine-learning technique is reinitialized for each trial.

\item After all the trials, the evaluation metrics are averaged 
to assess the performance of the machine-learning techniques.
\end{enumerate}

The average evaluation metrics for the machine-learning techniques
are shown in Table~\ref{Table:ResultSummary}. 
The experiment results show the following:

\begin{itemize}
\item The \textit{event classifier transformer} trained with DisCo loss shows
the highest significance with the lowest background correlation between
the machine-learning techniques.

\item The deep feed-forward network and \textit{event classifier transformer}
trained with BCE loss show the highest AUC and significance 
but have higher background correlations.

\item The neural networks trained with the DisCo loss show the lowest background
correlation.

\item Networks trained with \textit{extreme} + DisCo loss are trained 
more quickly compared to networks trained with BCE + Disco loss,
while showing similar performance.
\end{itemize}

\begin{table}[bh]
\caption{Average evaluation metrics for machine-learning techniques using the
random subsampling evaluation procedure with three trials. 
Random is a random classifier.
FNN is a feed-forward network.
DFNN is a deep feed-forward network. 
ETN is an \textit{event classifier transformer} network. 
N.A. means not available.
Ext is \textit{extreme} loss. 
Signi. is the expected significance.
AUC is the area under the curve of the receiver operating characteristic curve.
Bkg. Corr. is the correlation
of $m_{\ell\ell\gamma}$ with the machine-learning technique output
for the background calculated with the method described in Sec. \ref{Sec:EvaluationMetrics}. 
Best epoch is the epoch that had the highest significance on the validation dataset.
Bold numbers indicate the best
values over the machine-learning techniques.
\label{Table:ResultSummary}}
\begin{ruledtabular}
\begin{tabular}{cccccc}
\multirow{2}{*}{Technique} & \multirow{2}{*}{Loss} & \multirow{2}{*}{Signi.} & \multirow{2}{*}{AUC (\%)} & Bkg. Corr. & Best\tabularnewline
 & &  &  & ($10^{-3}$) &  epoch\tabularnewline
\hline
Random  & N.A       & 0.50 & 50.0 & 0   & N.A.\tabularnewline
BDT     & N.A       & 1.17 & 75.6 & 3.3 & N.A.\tabularnewline
\vspace{4pt}
XGBoost & N.A       & 1.49 & 75.8 & 2.7 & N.A.\tabularnewline
FNN     & BCE       & 1.43 & 75.2 & 2.8 & 22\tabularnewline
FNN     & Ext       & 1.45 & 75.4 & 2.7 & 107\tabularnewline
DFNN    & BCE       & \textbf{1.50} & \textbf{76.0} & 2.0 & 907\tabularnewline
DFNN    & Ext       & 1.46 & 75.7 & 2.0 & 487\tabularnewline
ETN     & BCE       & \textbf{1.51} & \textbf{76.2} & 2.3 & 570\tabularnewline
\vspace{4pt}
ETN     & Ext       & 1.48 & 75.7 & 2.3 & 307\tabularnewline
FNN     & BCE+DisCo & 1.35 & 75.0 & \textbf{0.7} & 633\tabularnewline
FNN     & Ext+DisCo & 1.46 & 74.7 & \textbf{0.8} & 327\tabularnewline
DFNN    & BCE+DisCo & 1.46 & 75.1 & \textbf{1.0} & 550\tabularnewline
DFNN    & Ext+DisCo & 1.47 & 75.7 & \textbf{1.0} & 273\tabularnewline
ETN     & BCE+DisCo & \textbf{1.52} & 75.6 & \textbf{1.0} & 670\tabularnewline
ETN     & Ext+DisCo & \textbf{1.50} & 75.4 & \textbf{1.0} & 623\tabularnewline
\end{tabular}
\end{ruledtabular}
\end{table}

There is an upper limit on performance for a classifier 
due to the similarities of the signal and background,
where machine-learning classifiers could be 
optimized in approaching the upper limit in a certain phase spaces.
Therefore the goal for the machine-learning classifier is 
to approach the upper limit in the phase space that is most relevant for the problem.
The \textit{event classifier transformer} tends to be able to approach the upper limit on the significance metric
better than the other machine-learning techniques used in this paper.
However with more hypertraining on the other machine-learning techniques or by using different machine-learning techniques,
it could be possible to get closer to the upper limit, which is left for future studies.

\begin{figure}
\includegraphics[width=0.45\textwidth]{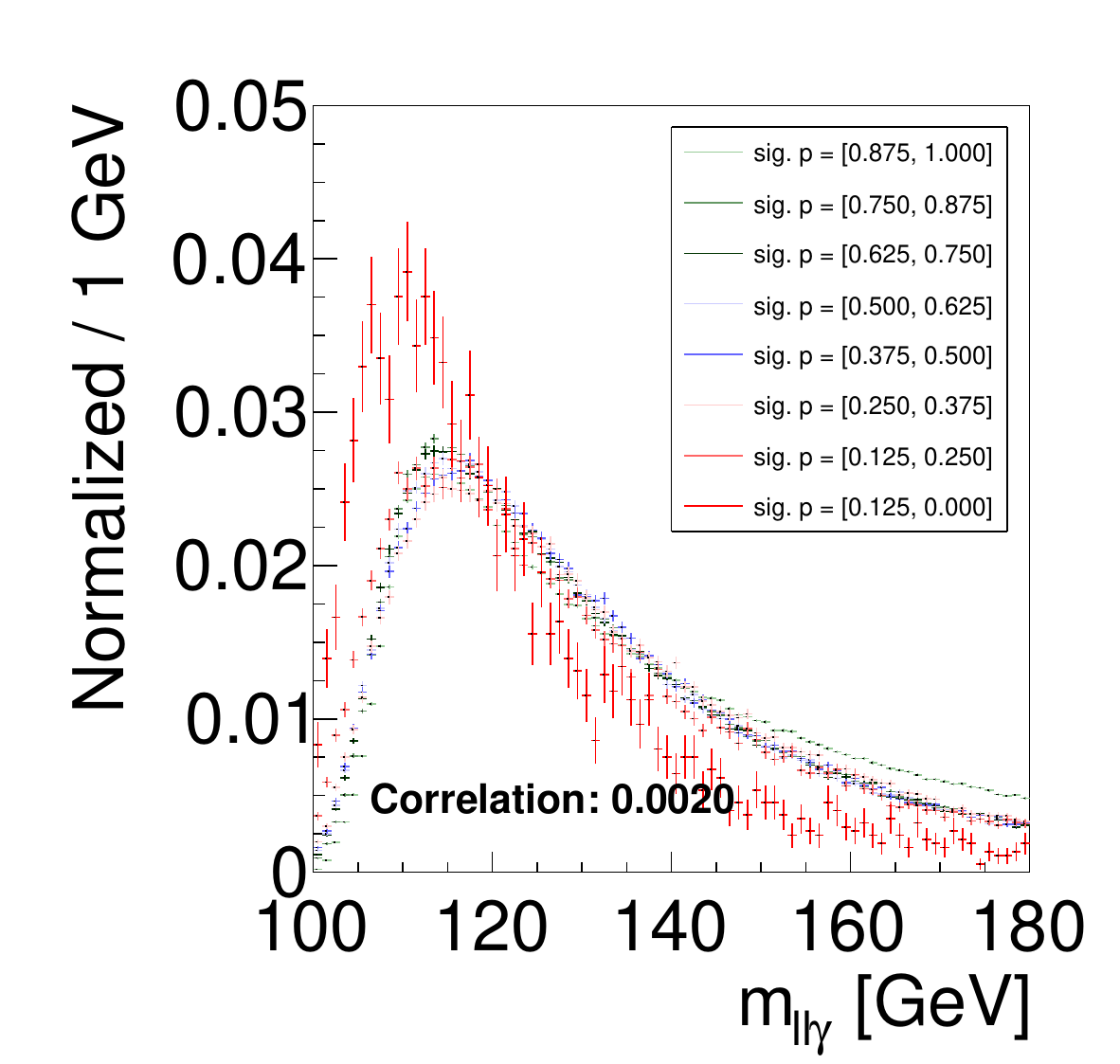}\\
\includegraphics[width=0.45\textwidth]{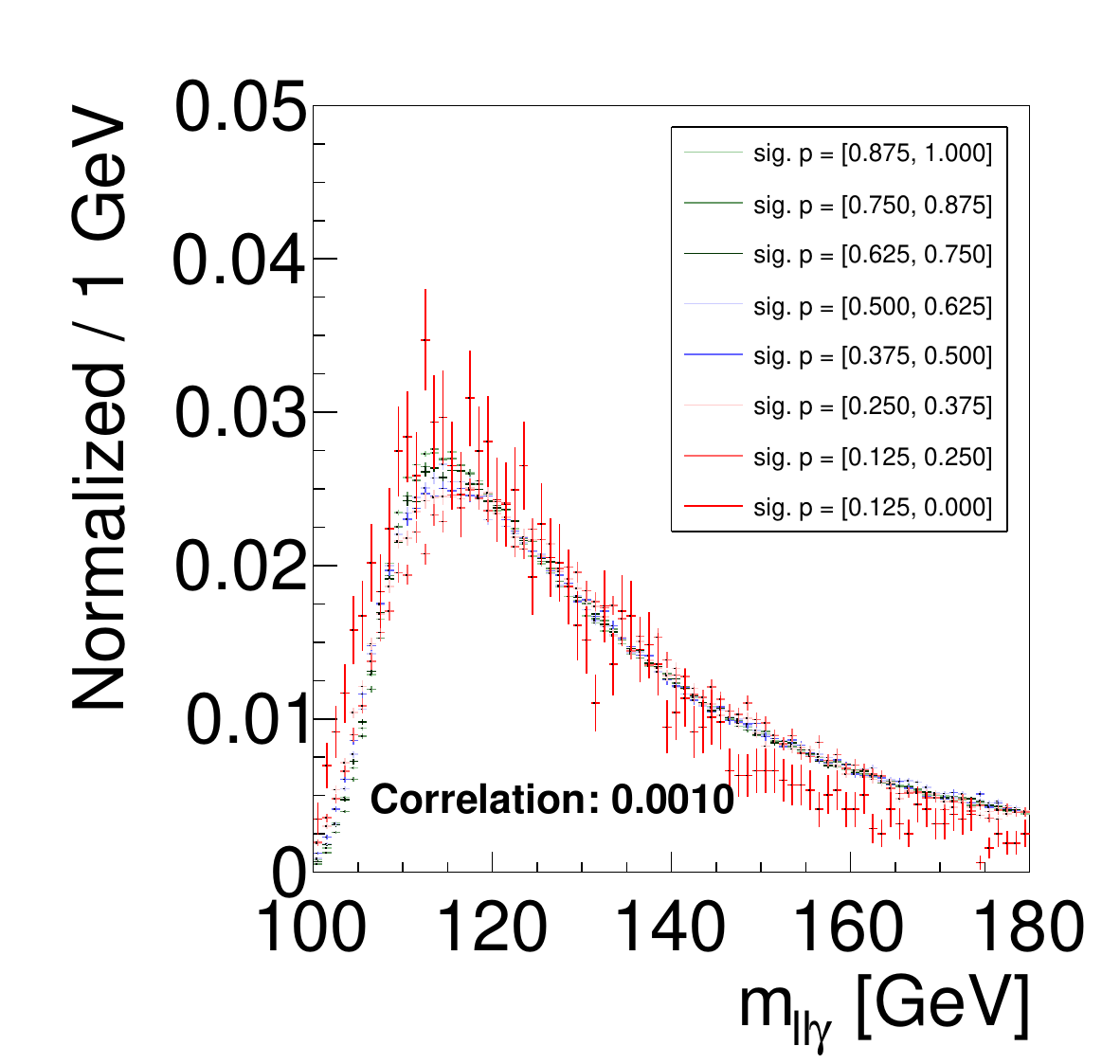}
\caption{$m_{\ell\ell\gamma}$ distribution of the background, where each histogram
is a bin in the machine-learning technique output distribution. 
Each bin has an equal number of signal events. 
Lower signal percentile (sig. p) values correspond to higher network output values. 
Top: deep feed-forward network trained with BCE loss. 
Bottom: \textit{Event classifier transformer} network trained with \textit{extreme} + DisCo loss. 
Correlation represents the magnitude of difference in the shapes between the machine-learning bins.
A lower correlation can be observed with the network trained with DisCo loss. 
\label{fig:mllg_correlation}}
\end{figure}

\begin{figure*}
\includegraphics[width=0.45\textwidth]{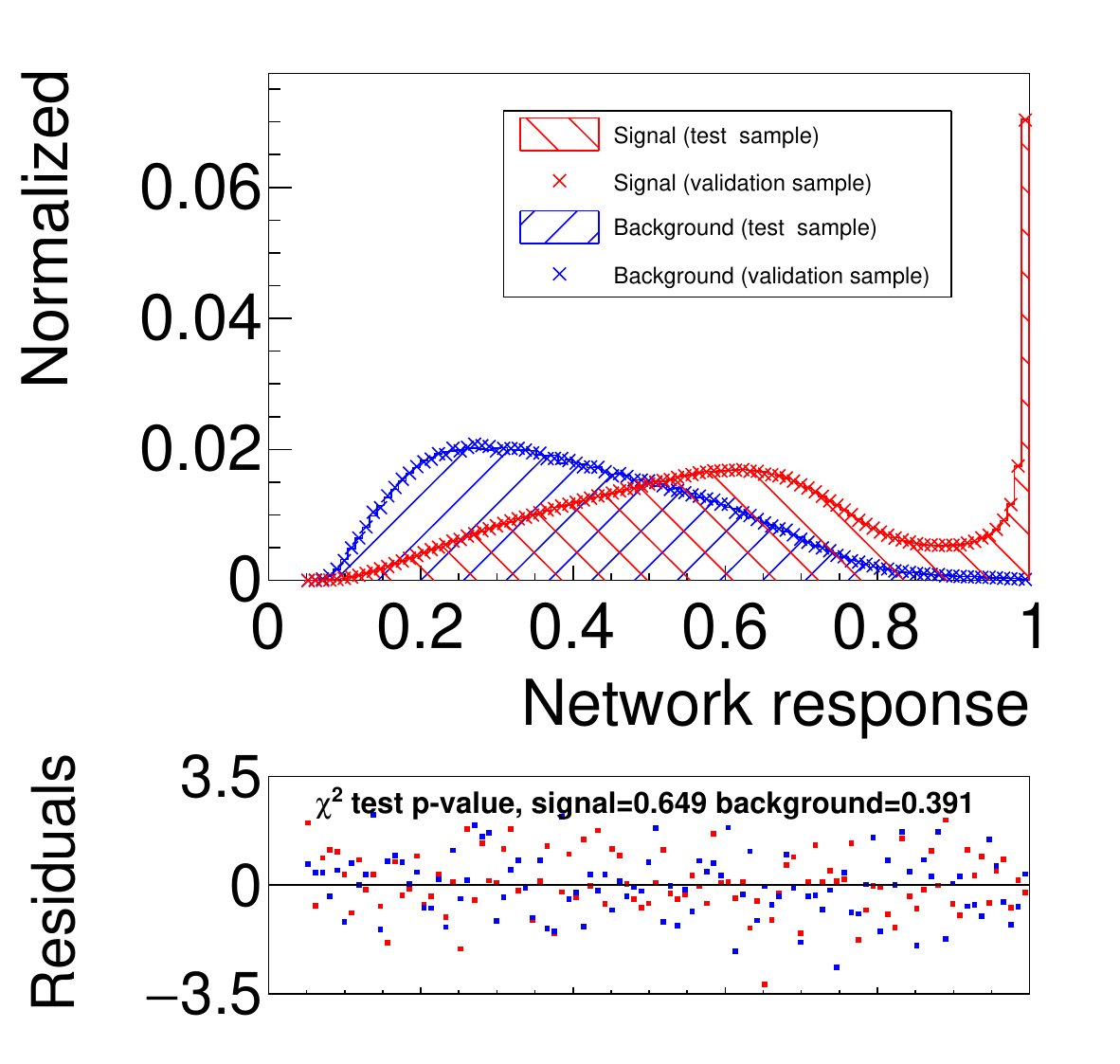}\includegraphics[width=0.45\textwidth]{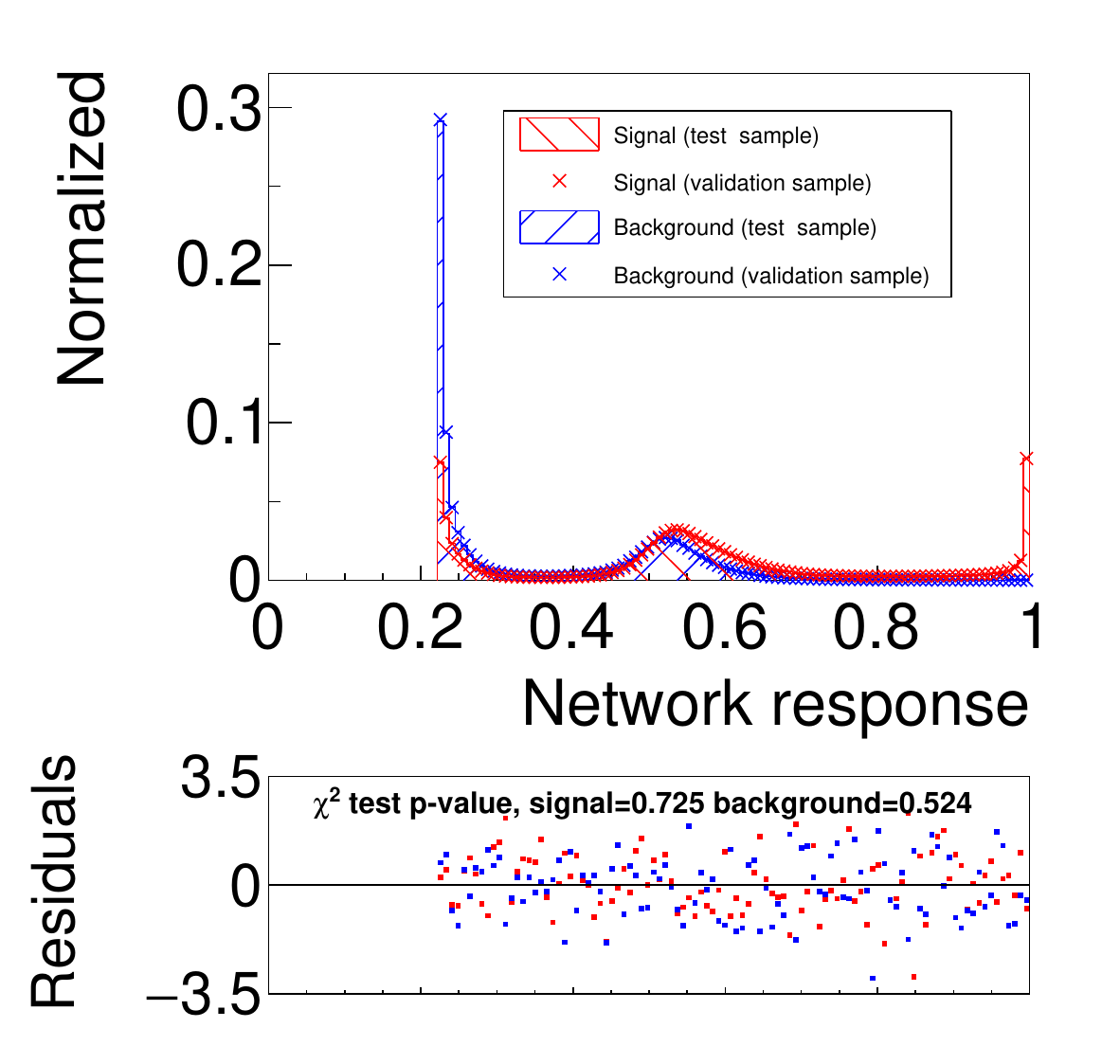}
\caption{
Left: deep feed-forward network trained with BCE loss. 
Right: \textit{event classifier transformer} network trained with the \textit{extreme} + DisCo loss. 
Top: network output distribution on the training dataset and validation dataset. 
Bottom: overtraining of the network evaluated by comparing the network output distribution between
the training and validation datasets for the signal and background using a $\chi^{2}$~test,
where residuals are the normalized differences between the output distributions.
\label{fig:overtraining}}
\end{figure*}

The machine-learning technique metrics
are shown for one of the subsampling trials. 
The correlation between the $m_{\ell\ell\gamma}$
and network output is shown in Fig.~\ref{fig:mllg_correlation}
for the deep feed-forward network trained with BCE loss 
and the \textit{event classifier transformer}
trained with the \textit{extreme} + DisCo loss.
The network output distribution is shown in Fig.~\ref{fig:overtraining},
where in the figure overtraining of the network is measured
with the $\chi^{2}$~test \cite{Chi2Test} implemented in ROOT \cite{ROOT1},
by comparing the network output distributions on the training sample and validation sample.
The residuals in the plot are the normalized differences between the output distributions
defined in \cite{Chi2Test}.
For both networks, the bump in the middle of the signal distribution is related with the networks 
making different output distributions depending on
$p_{T}^{\ell\ell\gamma}/m_{\ell\ell\gamma}$. 
An effect from the DisCo term can be seen, where the minimum network output value is shifted upwards. 

It is noted that when increasing the number of input features,
the increase in the number of trainable network weights 
for the \textit{event classifier transformer}
is small compared to a deep feed-forward network.
The smaller increase can make the network easier to train,
when more input features are used.
For example, when changing the number of input features 
from 12 to 15, 
the increase of the number of trainable weights 
for the \textit{event classifier transformer} is 265 (a 5\% increase),
while for the deep feed-forward neural network it is 4,671 (a 55\% increase).

\begin{figure}
\begin{centering}
\includegraphics[width=0.25\textwidth]{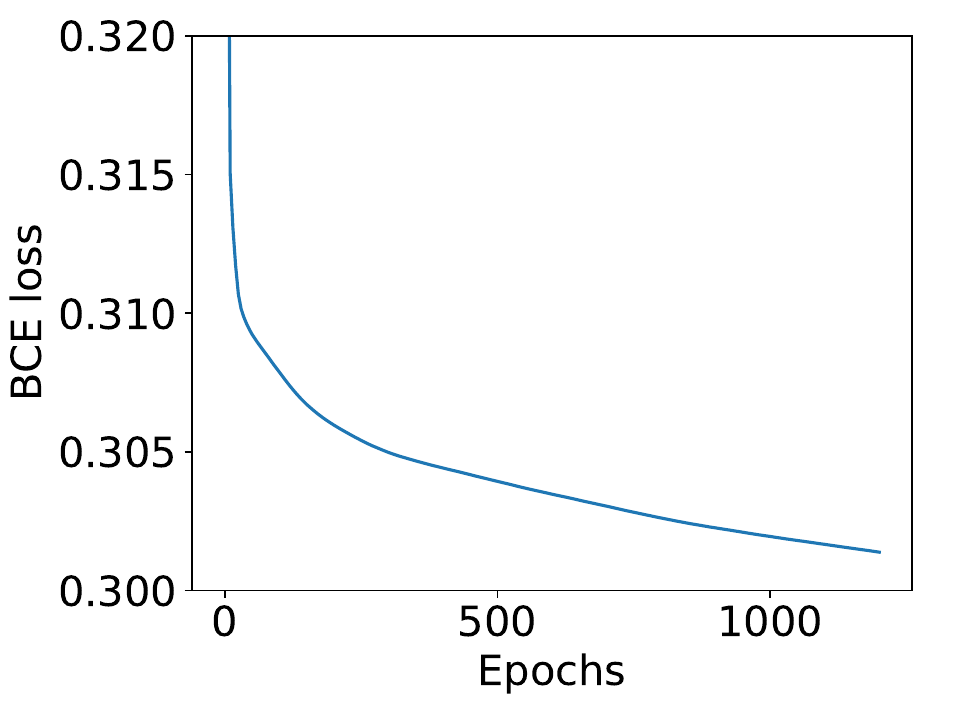}\includegraphics[width=0.25\textwidth]{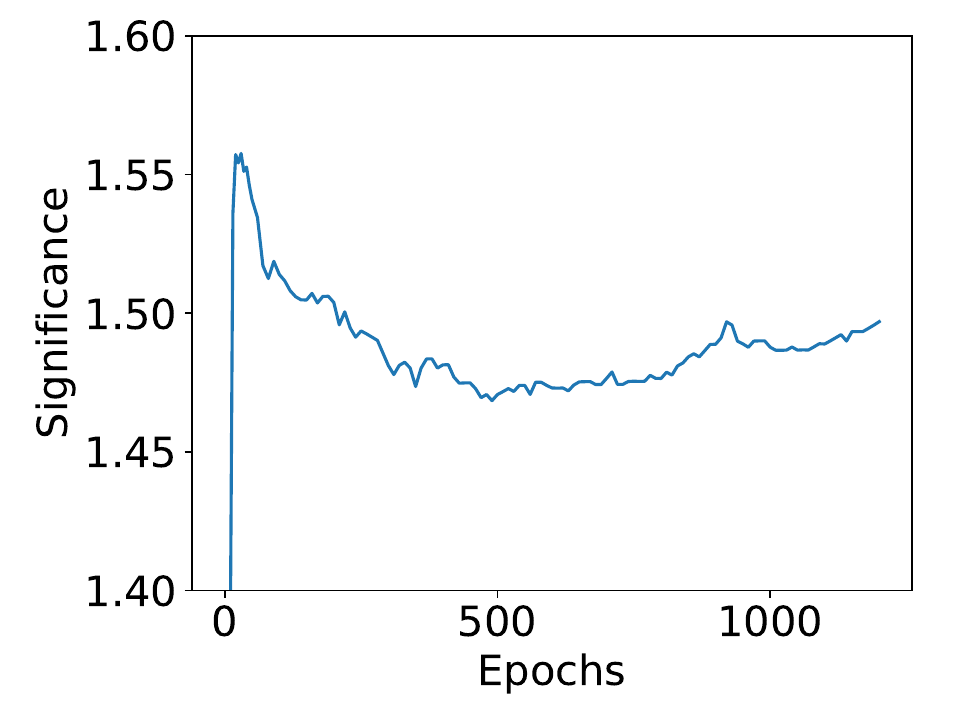}
\par\end{centering}
\caption{BCE loss (left) and expected significance (right) over training epochs for a feed-forward
neural network trained with BCE loss. 
The epoch that has the lowest BCE loss does not correspond 
to the epoch that has the highest significance.
The network has a high significance in the beginning of the training, 
and it becomes lower as the training continues. 
This behavior is due to the neural network becoming more discriminative for neural network
outputs at medium values, but as a tradeoff, it becomes less discriminative
for neural network outputs at high values, where the significance dependence is high.
\label{fig:lossVsSignificance}}
\end{figure}

During training of the feed-forward network
with BCE loss, the BCE loss value and the significance can have different
trends over the training epochs, 
which is shown in Fig.~\ref{fig:lossVsSignificance}.
This discrepancy demonstrates that when selecting the best model between the epochs,
the expected significance should be used instead of the loss.

A study was also performed to compare the results 
when the \textit{data scope training} method was not applied.
Instead, the wide mass window was used for both the classifier loss and the DisCo loss.
The training became difficult where the correlation could be high
or the network outputs could converge to one value.
The networks that were successfully trained had a few-percent poorer significance 
when having similar background correlations.

\section{Related work}

The Particle Transformer \cite{ParticleTransformer} is a transformer-based
neural network targeted toward jet-flavor tagging. 
The structure is similar to the transformer \cite{Transformer}, 
where each particle in a jet is considered as a token. 
Additionally, variables calculated using 
the features of a pair of particles are 
passed through a feed-forward network 
and added to the attention scores of the MHA. 
For the final prediction, 
class tokens are passed as a query to a MHA.
The \textit{event classifier transformer} in this work 
has a similar structure where the main difference is that 
each event feature is embedded to a separate token 
using separate feed-forward networks
and that the neural network output
is used to bin a bump hunt analysis.

There has been work to develop a loss function targeted toward significance,
where a loss based on the inverse of the significance \cite{SignificanceLoss} 
has been studied. However when a network is trained with this loss, 
the neural network outputs are clustered around 0 or 1.
This behavior makes the loss unusable for binning a bump hunt analysis. 
The proposed \textit{extreme} loss can enhance the significance
while not having the network outputs clustered around 0 or 1. 
It can also reduce the number of epochs that are needed 
to reach the optimal performance of the network.

DisCo \cite{DisCo} has been used for jet flavor tagging and for the
ABCD analysis technique \cite{DiscoABCD}, where it has been shown
to be effective in decorrelating a targeted feature of the jet with
the neural network output and decorrelating two neural network outputs
used for the ABCD method. 
In this work, DisCo is used to decorrelate the neural network output 
with the reconstructed mass to bin a bump hunting analysis.

\section{Summary and conclusions}

A transformer-based neural network is proposed to increase the expected significance 
of a search for a resonance in a reconstructed mass distribution.
The significance is enhanced by binning events 
using a network that discriminates between signal and background events.
To apply the transformer architecture for this task,
each event feature is passed through a separate feed-forward network
to create tokens for the transformer.
This network is called the \textit{event classifier transformer}.

Special training techniques are proposed to enhance the significance 
and reduce the correlation between the network's output and reconstructed mass.
\begin{itemize}
\item \textit{Extreme} loss is proposed that can enhance the significance 
and reduce the number of training epochs 
compared to the commonly used binary cross-entropy loss.

\item DisCo can be used to reduce the correlation.
This allows the network to have input event features
that are correlated with the reconstructed mass.

\item \textit{Data scope training} is proposed,
where loss terms have different data scopes.
This method can increase the significance and reduce the correlation.

\item A \textit{significance selection metric} is proposed for choosing the best model between the training epochs
instead of loss.
\end{itemize}

In the context of a simplified $H\rightarrow Z\left(\ell^{+}\ell^{-}\right)\gamma$ search,
the new \textit{event classifier transformer} trained with the special techniques
shows higher significance and lower mass correlation 
when compared with boosted decision trees and feed-forward networks.
This result demonstrates the potential of the \textit{event classifier transformer}
and the specialized training techniques 
targeted toward binning a search for a resonance in the reconstructed mass distribution.

\begin{acknowledgments}
I am grateful to Gyuwan Kim in the UCSB Computer Science 
and Natural Language Processing group
for introducing me to the transformer neural network and for revising the paper.
I am also grateful to Jeff Richman in the UCSB High Energy Physics group 
for his guidance and help in editing the paper. 
The material is based upon work supported by the U.S. Department of Energy, 
Office of Science, Office of High Energy Physics under Award Number DE-SC0011702.
\end{acknowledgments}

\bibliography{event_transformer}

\end{document}